%% file: nyt-kp-generation.tex
\documentclass[11pt,a4paper]{article}
\usepackage[hyperref]{acl2019}
\usepackage{times}
\usepackage{latexsym}
\usepackage{amsmath}
\usepackage{booktabs}
\usepackage{url}
\usepackage{graphicx}
\usepackage{tikz}
\usepackage{pgfplots}
\usepackage{subcaption}
\usetikzlibrary{patterns}
\pgfplotsset{compat=1.14}

\usepackage{multirow}

\newcommand{\dataset}{\textit{KPTimes}}
\newcommand{\jptimes}{\textit{JPTimes}}
\newcommand{\dataseturl}{\url{https://github.com/ygorg/KPTimes}}

\colorlet{bestmodel}{blue!5}

\newcommand\best[2][n]{%
  \ifx s#1\colorbox{bestmodel}{#2}\hspace{-.3em}$^\dagger$\else
  \ifx n#1\colorbox{bestmodel}{#2}\else
  \ifx x#1\fbox{#2}\hspace{-.3em}$^\dagger$\else
  \mathrm{Illegal~option}%
  \fi\fi\fi
}

\aclfinalcopy 

\title{\dataset: A Large-Scale Dataset for Keyphrase Generation on News Documents}

\author{Ygor Gallina \\\And
  Florian Boudin \\\\
  LS2N, Universit\'e de Nantes, France\\
  {\tt firstname.lastname@univ-nantes.fr} \\\And
  Beatrice Daille \\}

\date{}

\begin{document}
\maketitle
\begin{abstract}
Keyphrase generation is the task of predicting a set of lexical units that conveys the main content of a source text.
Existing datasets for keyphrase generation are only readily available for the scholarly domain and include non-expert annotations.
In this paper we present \dataset{}, a large-scale dataset of news texts paired with editor-curated keyphrases.
Exploring the dataset, we show how editors tag documents, and how their annotations differ from those found in existing datasets.
We also train and evaluate state-of-the-art neural keyphrase generation models on \dataset{} to gain insights on how well they perform on the news domain. 
The dataset is available online at \dataseturl{}.
\end{abstract}

\section{Introduction}

\input{tables/example.tex}

Keyphrases are single or multi-word lexical units that best summarise a document~\cite{evans-zhai:1996:ACL}.
As such, they are of great importance for indexing, categorising and browsing digital libraries~\cite{witten2009build}.
Yet, very few documents have keyphrases assigned, thus raising the need for automatic keyphrase generation systems.
This task falls under the task of automatic keyphrase extraction which can also be the subtask of finding keyphrases that only appear in the input document.
Generating keyphrases can be seen as a particular instantiation of text summarization, where the goal is not to produce a well-formed piece of text, but a coherent set of phrases that convey the most salient information.
Those phrases may or may not appear in the document, the latter requiring some form of abstraction to be generated.
State-of-the-art systems for this task rely on recurrent neural networks~\cite{P17-1054,chen-EtAl:2018:EMNLP9,chen-etal-2019-integrated}, and hence require large amounts of annotated training data to achieve good performance.
As gold annotated data is expensive and difficult to obtain~\cite{Mao2017}, previous works focused on readily available scientific abstracts and used author-assigned keyphrases as a proxy for expert annotations.
However, this poses two major issues:
1)~neural models for keyphrase generation do not generalize well across domains, thus limiting their use in practice;
2)~author-assigned keyphrases exhibit strong consistency issues that negatively impacts the model's performance.
There is therefore a great need for annotated data from different sources, that is both sufficiently large to support the training of neural-based models and that comprises gold-standard labels provided by experts.
In this study, we address this need by providing \dataset{}, a dataset made of 279\,923 news articles that comes with editor-assigned keyphrases.

Online news are particularly relevant to keyphrase generation since they are a natural fit for faceted navigation~\cite{tunkelang2009faceted} or topic detection and tracking~\cite{allan2012topic}.
Also, and not less importantly, they are available in large quantities and are sometimes accompanied by metadata containing human-assigned keyphrases initially intended for search engines.
Here, we divert these annotations from their primary purpose, and use them as gold-standard labels to automatically build our dataset.
More precisely, we collect data by crawling selected news websites and use heuristics to draw texts paired with gold keyphrases. 
We then explore the resulting dataset to better understand how editors tag documents, and how these expert annotations differ from author-assigned keyphrases found in scholarly documents.
Finally, we analyse the performance of state-of-the-art keyphrase generation models and investigate their transferability to the news domain and the impact of domain shift.

\section{Existing datasets}

Frequently used datasets for keyphrase generation have a common characteristic that they are, by and large, made from scholarly documents (abstracts or full texts) paired with non-expert (mostly from authors) annotations.
Notable examples of such datasets are SemEval-2010~\cite{kim-EtAl:2010:SemEval} and KP20k~\cite{P17-1054}, which respectively comprises scientific articles and paper abstracts, both about computer science and information technology.
Detailed statistics are listed in Table~\ref{tab:stats}.
Only two publicly available datasets, that we are aware of, contain news documents: DUC-2001~\cite{Wan:2008:SDK:1620163.1620205} and KPCrowd~\cite{MARUJO12.672}.
Originally created for the DUC evaluation campaign on text summarization~\cite{overintroduction}, the former is composed of 308 news annotated by graduate students.
The latter includes 500 news annotated by crowdsourcing.
Both datasets are very small and contain newswire articles from various online sources labelled by non-expert annotators, in this case readers, which is not without issues.

\input{tables/datasets_statistics.tex}

Thus, unlike author annotations, those produced by readers exhibit significantly lower missing keyphrases, that is, gold keyphrases that do not occur in the content of the document.
In the DUC-2001 dataset for example, more than 96\% of the gold keyphrases actually appear in the documents.
This confirms previous observations that readers tend to assign keyphrases in an extractive fashion~\cite{10.1007/978-3-319-19548-3_21}, which makes these datasets less suitable for the task at hand (keyphrase generation) but rather relevant for a purely extractive task (keyphrase extraction).
Yet, author-assigned keyphrases commonly found in scientific paper datasets are not perfect either, as they are less constrained~\cite{DBLP:conf/icwsm/SoodOHB07} and include seldom-used variants or misspellings that negatively impact performance.
One can see there is an apparent lack of sizeable expert-annotated data that enables the development of neural keyphrase generation models in a domain other than scholarly texts.
Here, we fill this gap and propose a large-scale dataset that includes news texts paired with manually curated gold standard annotations.

\section{Building the \dataset{} dataset}
\label{sec:building}

To create the \dataset{} dataset, we collected over half a million newswire articles by crawling selected online news websites.
We applied heuristics to identify the content (title, headline and body) of each article and regarded the keyphrases provided in the HTML metadata as the gold standard.
A cherry-picked sample document is showcased in Figure~\ref{fig:example}, it allows to show present and absent keyphrases, as well as keyphrase variants (in this example \texttt{News media} and \texttt{journalism}).

We use the New York Times\footnote{\url{https://www.nytimes.com/}} as our primary source of data, since the content tagging policy that it applies is rigorous and well-documented\footnote{\url{https://lac-group.com/rules-based-tagging-metadata/}}.
The news articles are annotated in a semi-automatic way, first the editors revise a set of tags proposed by an algorithm. They then provide additional tags which will be used by a taxonomy team to improve the algorithm.

We first retrieved the URLs of the free-to-read articles from 2006 to 2017\footnote{\url{https://spiderbites.nytimes.com/}}, and collected the corresponding archived HTML pages using the Internet Archive\footnote{\url{https://archive.org/}}.
Doing so allows the distribution of our dataset using a thin, URL-only list.
We then extracted the HTML body content using \texttt{beautifulsoup}\footnote{\url{https://www.crummy.com/software/BeautifulSoup/}} and devised heuristics to extract the main content and title of each article while excluding extraneous HTML markup and inline ads.
Gold standard keyphrases are obtained from the metadata (field types \texttt{news\_keywords} and \texttt{keywords}\footnote{The change of field name correspond to the introduction of the \emph{keywords} tag as a W3C standard.}) available in the HTML page of each article.
Surface form variants of gold keyphrases (e.g.~``\textit{AIDS}; \textit{HIV}'', ``\textit{Driverless Cars}; \textit{Self-Driving Cars}'' or ``\textit{Fatalities}; \textit{Casualties}''), which are sometimes present in the metadata, are kept to be used for evaluation purposes. 

We further cleansed and filtered the dataset by removing duplicates, articles without content and those with too few (less than 2) or too many (more than 10) keyphrases.
This process resulted in a set of 279\,923 article-keyphrase pairs.
We randomly divided this dataset into training (92.8\%), development (3.6\%) and test  (3.6\%) splits.

Restricting ourselves to one source of data ensures the uniformity and consistency of annotation that is missing in the other datasets, but it may also make the trained model source-dependent and harm generalization.
To monitor the model's ability to generalize, we gather a secondary source of data.
We collected HTML pages from the Japan Times\footnote{\url{https://www.japantimes.co.jp/}} and processed them the same way as described above.
10K more news articles were gathered as the \jptimes{} dataset.

Although in this study we concentrate only on the textual content of the news articles, it is worth noting that the HTML pages also provide additional information that can be helpful in generating keyphrases such as text style properties (e.g.~bold, italic), links to related articles, or news categorization (e.g.~politics, science, technology).

\section{Data analysis}

We explored the \dataset{} dataset to better understand how it stands out from the existing ones.
First, we looked at how editors tag news articles.
Figure~\ref{fig:distrib} illustrates the difference between the annotation behaviour of readers, authors and editors through the number of times that each unique keyphrase is used in the gold standard.
We see that non-expert annotators use a larger, less controlled indexing vocabulary, in part because they lack the higher level of domain expertise that editors have.
For example, we observe that frequent keyphrases in \dataset{} are close to topic descriptors (e.g.~``\textit{Baseball}``, ``\textit{Politics and Government}``) while those appearing only once are very precise (e.g.~``\textit{Marley's Cafe}``, ``\textit{Catherine E. Connelly}``).
Annotations in \dataset{} are arguably more uniform and consistent, through the use of tag suggestions, which, as we will soon discuss in \S\ref{sec:results}, makes it easier for supervised approaches to learn a good model.

\input{figures/distributions.tex}

Next, we further looked at the characteristics of the gold keyphrases in \dataset{}.
Table~\ref{tab:stats} shows that the number of gold keyphrases per document is similar to the one observed for KP20k while the number of missing keyphrases is higher.
This indicates that editors are more likely to generalize and assign keyphrases that do not occur in the document ($\approx 55\%$).
It is therefore this ability to generalize that models should mimic in order to perform well on \dataset{}.
We also note that keyphrases are on average shorter in news datasets ($1.5$ words) than those in scientific paper datasets ($2.4$ words).
This may be due to the abundant use of longer, more specific phrases in scholarly documents~\cite{jin-etal-2013-mining}.

Variants of keyphrases recovered from the metadata occur in 8\% of the documents and represent 810 sets of variants in the \dataset{} test split.
These variants often refer to the same concept (e.g.~``\textit{Marijuana; Pot; Weed}``), but can sometimes be simply semantically related (e.g.~``\textit{Bridges; Tunnels}``).
Thereafter, keyphrase variants will be used during model evaluation for reducing the number of mismatches associated with commonly used lexical overlap metrics.

\section{Performance of existing models}

We train and evaluate several keyphrase generation models to understand the challenges of \dataset{} and its usefulness for training models.

\input{tables/performance.tex}

\subsection{Evaluation metrics}

We follow the common practice and evaluate the performance of each model in terms of f-measure (F$_1$) at the top $N=10$ keyphrases, and apply stemming to reduce the number of mismatches.
We also report the Mean Average Precision (MAP) scores of the ranked lists of keyphrases.

\subsection{Models}

\subsubsection*{Baseline: FirstPhrase}

Position is a strong feature for keyphrase extraction, simply because texts are usually written so that the most important ideas go first~\cite{marcu1997rhetorical}.
In news summarization for example, the lead baseline --that is, the first sentences from the document--, while incredibly simple, is still a competitive baseline~\cite{kedzie-mckeown-daumeiii:2018:EMNLP}.
Similar to the lead baseline, we compute the \textbf{FirstPhrases} baseline that extracts the first $N$ keyphrase candidates\footnote{Sequences of adjacent nouns with one or more preceding adjectives of length up to five words.} from a document.

\subsubsection*{Baseline, unsupervised: MultipartiteRank}

The second baseline we consider, \textbf{MultipartiteRank}~\cite{boudin:2018:N18-2}, represents the state-of-the-art in unsupervised graph-based keyphrase extraction.
It relies on a multipartite graph representation to enforce topical diversity while ranking keyphrase candidates.
Just as FirstPhrases, this model is bound to the content of the document and cannot generate missing keyphrases.
We use the implementation of MultipartiteRank available in \texttt{pke}\footnote{\url{https://github.com/boudinfl/pke}}~\cite{boudin-2016-pke}.

\subsubsection*{State-of-the-art, supervised: CopyRNN}

The generative neural model we include in this study is \textbf{CopyRNN}~\cite{P17-1054}, an encoder-decoder model that incorporates a copying mechanism~\cite{gu:2016:ACL} in order to be able to generate phrases that rarely  occur.
When properly trained, this model was shown to be very effective in extracting keyphrases from scientific abstracts.
CopyRNN has been further extended by~\cite{chen-EtAl:2018:EMNLP9} to include correlation constraints among keyphrases which we do not include here as it yields comparable results.

Two models were trained to bring evidence on the necessity to have datasets from multiple domains. CopySci was trained using scientific abstracts (KP20k) and CopyNews using newspaper articles (\dataset{}), the two models use the same architecture.

\subsection{Results}
\label{sec:results}

Model performances for each dataset are reported in Table~\ref{tab:performance}.
Extractive baselines show the best results for KPCrowd and DUC-2001 which is not surprising given that these datasets exhibit the lowest ratio of absent keyphrases.
Neural-based models obtain the greatest performance, but only for the dataset on which they were trained.
We therefore see that these models do not generalize well across domains, confirming previous preliminary findings~\cite{P17-1054} and exacerbating the need for further research on this topic.
Interestingly, CopyNews outperforms the other models on \jptimes{} and achieves very low scores for KPCrowd and DUC-2001, although all these datasets are from the same domain.
This emphasizes the differences that exist between the reader- and editor-assigned gold standard.
The score difference may be explained by the ratio of absent keyphrases that differs greatly between the reader-annotated datasets and \jptimes{} (see Table~\ref{tab:stats}), and thus question the use of these rather extractive datasets for evaluating keyphrase generation.

Finally, we note that the performance of CopyNews on \dataset{} is significantly higher than that of CopySci on KP20k, proving that a more uniform and consistent annotation makes it easier to learn a good model.

\section{Conclusion}

In this paper we presented \dataset{}, a large-scale dataset of newswire articles to train and test deep learning models for keyphrase generation.
The dataset and the code are available at \dataseturl{}.
Large datasets have driven rapid improvement in other natural language generation tasks, such as machine translation or summarization.
We hope that \dataset{} will play this role and help the community in devising more robust and generalizable neural keyphrase generation models.

\bibliography{biblio}
\bibliographystyle{acl_natbib}

\end{document}

%% file: tables/example.tex
\definecolor{yellow}{rgb}{0.99, 0.76, 0.0} 
\definecolor{green}{rgb}{0.0, 0.62, 0.38} 
\definecolor{blue}{rgb}{0.13, 0.67, 0.8} 
\definecolor{red}{rgb}{0.87, 0.36, 0.51} 
\definecolor{orange}{rgb}{1.0, 0.43, 0.29} 

\begin{figure*}[!htbp]
    \centering
    \resizebox{0.95\textwidth}{!}{%
    \begin{tabular}{|p{0.98\textwidth}|}
    \textbf{\textcolor{yellow}{Muslim} \textcolor{orange}{Women} in Hijab Break Barriers: \lq{}Take the Good With the Bad\rq{}}
    
    \vspace{0.5em}
    
    When Ginella Massa, a Toronto-based TV reporter, recently accepted a request to host an evening newscast, she was not planning or expecting to make history for wearing a hijab. She was just covering for a colleague who wanted to go to a hockey game. And that's how Ms.~Massa, who works at CityNews in Toronto, became the first Canadian woman to host a newscast from a large \textcolor{blue}{media} company while wearing the head scarf.
    [...]
    This new trend of inclusion occurs amid a more sinister one, as reported \textcolor{red}{hate crimes} against \textcolor{yellow}{Muslims} are on the rise in the United States and \textcolor{green}{Canada}. The F.B.I. says that a surge in \textcolor{red}{hate crimes} against \textcolor{yellow}{Muslims} has led to an overall increase in \textcolor{red}{hate crimes} in the United States; \textcolor{yellow}{Muslims} have borne the brunt of the increase with 257 recorded attacks.
    [...]
    In \textcolor{green}{Canada}, where Ms.~Massa has lived since she was a year old, the number of reported \textcolor{red}{hate crimes} has dropped slightly overall, but the number of recorded attacks against \textcolor{yellow}{Muslims} has grown: 99 attacks were reported in 2014, according to an analysis by the \textcolor{blue}{news} site Global \textcolor{blue}{News} of data from Statistics \textcolor{green}{Canada}, a government agency.
    [...]

    \vspace{0.5em}
    
    \textbf{keywords:} US; Islam; Fashion; \textcolor{yellow}{Muslim}~Veiling; \textcolor{orange}{Women} and Girls; (\textcolor{blue}{News media}, journalism); \textcolor{red}{Hate~crime}; \textcolor{green}{Canada}
    \end{tabular}%
    }
    \caption{Sample document from \dataset{} (id: 0296216). Keyphrases (or part of) appearing in the document are colored.}
    \label{fig:example}
\end{figure*}

\definecolor{green}{rgb}{0.0, 1.0, 0.0} 
\definecolor{blue}{rgb}{0.0, 0.0, 1.0} 
\definecolor{red}{rgb}{1.0, 0.0, 0.0} 

%% file: tables/datasets_statistics.tex
\begin{table*}[t]
    \centering
    \begin{tabular}{cr|crrrrrrr}
    \cmidrule[1pt]{2-10}
        &
        \textbf{Dataset} &
        \textbf{Ann.} &
        \textbf{\#Train} &
        \textbf{\#Dev} &
        \textbf{\#Test} &
        \textbf{\#words} &
        \textbf{\#kp} &
        \textbf{len kp} &
        \textbf{\%abs} \\
    \cmidrule{2-10}
    \multirow{2}{*}[+0.3ex]{\rotatebox{90}{\small{\textbf{Scholar}}}}
      & SemEval-2010          & $A \cup R$ & 144  & -   & 100 & 7\,961 & 14.7 & 2.2 & 19.7 \\
      & KP20k                 & $A$   & 530K & 20K & 20K & 176    & 5.3  & 2.6 & 42.6 \\
    \cmidrule{2-10}
    \multirow{4}{*}[0ex]{\rotatebox{90}{\small{\textbf{News}}}}
      & DUC-2001              & $R$   & -    & -   & 308 & 847    & 8.1  & 2.0 &  3.7 \\
      & KPCrowd               & $R$   & 450  & -   & 50  & 465    & 46.2 & 1.1 & 11.2 \\
      & \dataset{} (this work) & $E$   & 260K & 10K & 10K & 921    & 5.0  & 1.5  & 54.7 \\ 
      & \jptimes{} (this work)    & $A$   & -    & -   & 10K & 648    & 5.3  & 1.3 & 28.2 \\
    \cmidrule[1pt]{2-10}
    \end{tabular}%
    \caption{Statistics of available datasets for keyphrase generation.
    Gold annotation is performed by authors ($A$), readers ($R$) or editors ($E$).
    The number of documents in the training (\#Train), validation (dev) and testing (\#Test) splits are shown. The average number of keyphrases (\#kp) and words (\#words) per document, the average length of keyphrases (len kp) and the ratio of keyphrases in the reference that do not appear in the document (\%abs) are computed on the test set.}
    \label{tab:stats}
\end{table*}

%% file: figures/distributions.tex
\pgfplotsset{
    legend image code/.code={
        \draw [#1] (0cm,-0.1cm) rectangle (0.6cm,0.1cm);
    },
}
\usepgfplotslibrary{fillbetween}

\begin{figure}[ht!]
    \resizebox{0.48\textwidth}{!}{%
    \centering
    \begin{tikzpicture}
	    \begin{axis}[height=5cm,
	                 width=8.5cm,
	                 grid style={dashed,gray!30},
	                 ymajorgrids,
                     xlabel={Number of assignments},
                     ylabel={\% of gold keyphrases},
                     every node near coord/.append style={font=\footnotesize},
                     ybar=0pt,
                     bar width=9pt,
                     ymin=0,
                     ymax=100,
                     xmin=0.3,
                     xmax=5.7,
                     tick align=inside,
                     legend entries={KPCrowd (readers), KP20k (authors), \dataset{} (editors)},
                     legend cell align={left},
                     legend style={font=\small},
                     ]
                     
            
            
            \addplot [draw=green!80!black, fill=green!10] coordinates {
                (1, 86.8)
                (2, 9.0)
                (3, 2.9)
                (4, 1.0)
                (5, 0.1)
            };

            \addplot [draw=blue, fill=blue!15, pattern=north east lines] coordinates {
                (1, 79.4)
                (2, 10.2)
                (3, 3.5)
                (4, 1.8)
                (5, 1.1)
            };
            

            
            \addplot [draw=red, fill=red!15, nodes near coords, point meta=explicit symbolic] coordinates {
                (1, 64.4) 
                (2, 12.4) 
                (3, 6.1) 
                (4, 3.0) 
                (5, 2.0) 
            };
            
            \node (A1) at (axis cs:0.73,86.8) {};
            \node (A2) at (axis cs:1.26,64.4) {};
            \draw[->,>=latex,red!80!black] (A1) to[bend left] node (sum) [midway, right, font=\small] {\textcolor{red!80!black}{-22.4}} (A2);
            
            \node (B1) at (axis cs:1.73,9) {};
            \node (B2) at (axis cs:2.26,12.4) {};
            \draw[->,>=latex,green!60!black] (B1) to[bend left] node (sum) [midway, above, font=\small] {\textcolor{green!60!black}{+3.4}} (B2);
            
            \node (C1) at (axis cs:2.73,2.9) {};
            \node (C2) at (axis cs:3.26,6.1) {};
            \draw[->,>=latex,green!60!black] (C1) to[bend left] node (sum) [midway, above, font=\small] {\textcolor{green!60!black}{+3.2}} (C2);
            
            \node (D1) at (axis cs:3.73,1.0) {};
            \node (D2) at (axis cs:4.26,3.0) {};
            \draw[->,>=latex,green!60!black] (D1) to[bend left] node (sum) [midway, above, font=\small] {\textcolor{green!60!black}{+2.0}} (D2);
            
            \node (E1) at (axis cs:4.73,0.1) {};
            \node (E2) at (axis cs:5.26,2.0) {};
            \draw[->,>=latex,green!60!black] (E1) to[bend left] node (sum) [midway, above, font=\small] {\textcolor{green!60!black}{+1.9}} (E2);

        \end{axis}
    \end{tikzpicture}%
    }
    \vspace*{-1.5em}
    \caption{Distributions of  gold keyphrase assignments.}
    \label{fig:distrib}
    \vspace*{-1em}
\end{figure}
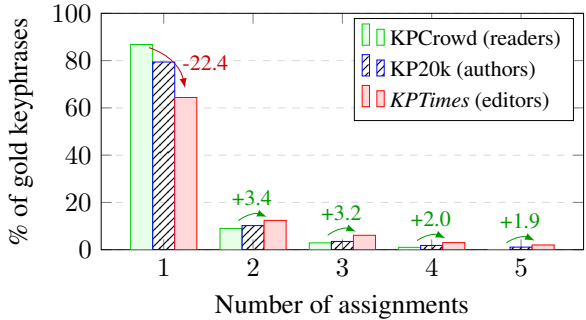

%% file: tables/performance.tex
\begin{table*}[ht!]
    \centering
    \resizebox{\textwidth}{!}{
    \begin{tabular}{r c@{\hspace*{1mm}}c  c@{\hspace*{1mm}}c  c@{\hspace*{1mm}}c c@{\hspace*{1mm}}c | c@{\hspace*{1mm}}c c@{\hspace*{1mm}}c}
        ~ & 
        \multicolumn{2}{c}{\textbf{KPCrowd}} &
        \multicolumn{2}{c}{\textbf{DUC}} &
        \multicolumn{2}{c}{\textbf{\textit{\dataset{}}}} &
        \multicolumn{2}{c}{\textbf{\textit{\jptimes{}}}} &
        \multicolumn{2}{c}{\textbf{SemEval}} &
        \multicolumn{2}{c}{\textbf{KP20k}} \\
        
        \cmidrule(lr){2-3} \cmidrule(lr){4-5} \cmidrule(lr){6-7} \cmidrule(lr){8-9} \cmidrule(lr){10-11} \cmidrule(lr){12-13} \\[-1.5em]
        
        ~ & \small{$\text{F}@10$} & \small{MAP} & \small{$\text{F}@10$} & \small{MAP} & \small{$\text{F}@10$} & \small{MAP} & \small{$\text{F}@10$} & \small{MAP} & \small{$\text{F}@10$} & \small{MAP} & \small{$\text{F}@10$} & \small{MAP} \\[-.2em]
        
        \midrule
        
        FirstPhrases & 17.1 & 16.5 & 24.6 & 22.3 & \phantom{0}9.2 & \phantom{0}8.4 & 13.5 & 13.1 & 13.8 & 10.5 & 13.5 & 12.6 \\
        MultipartiteRank & \best{18.2} & \best{17.0} & \best{25.6} & \best{24.9} & 11.2 & 10.1 & 16.9 & 16.5 & 14.3 & 10.6 & 13.6 & 13.3 \\
        CopySci & 15.5 & 11.1 & 12.7 & \phantom{0}9.7 & 11.0 & 10.6 & 18.9 & 19.8 & \best{20.3} & \best{13.8} & \best{25.4} & \best{28.7} \\
        CopyNews & \phantom{0}8.4 & \phantom{0}4.2 & 10.5 & \phantom{0}7.2 & \best{39.3} & \best{50.9} & \best{24.6} & \best{26.5} & \phantom{0}7.0 & \phantom{0}3.5 & \phantom{0}6.6 & \phantom{0}5.1\\

        \bottomrule
    \end{tabular}
    }
    \caption{Performance on benchmark datasets composed of newspaper article, full scientific article and scientific article abstract. The generation models CopySci and CopyNews were trained respectively on KP20k and \dataset{}. The dataset presented in this work are written in italic.}
    \label{tab:performance}
\end{table*}

%% file: nyt-kp-generation.bbl
\begin{thebibliography}{20}
\expandafter\ifx\csname natexlab\endcsname\relax\def\natexlab#1{#1}\fi

\bibitem[{Allan(2012)}]{allan2012topic}
James Allan. 2012.
\newblock \emph{Topic detection and tracking: event-based information
  organization}, volume~12.
\newblock Springer Science \& Business Media.

\bibitem[{Boudin(2016)}]{boudin-2016-pke}
Florian Boudin. 2016.
\newblock \href {https://www.aclweb.org/anthology/C16-2015} {pke: an open
  source python-based keyphrase extraction toolkit}.
\newblock In \emph{Proceedings of {COLING} 2016, the 26th International
  Conference on Computational Linguistics: System Demonstrations}, pages
  69--73, Osaka, Japan. The COLING 2016 Organizing Committee.

\bibitem[{Boudin(2018)}]{boudin:2018:N18-2}
Florian Boudin. 2018.
\newblock \href {http://www.aclweb.org/anthology/N18-2105} {Unsupervised
  keyphrase extraction with multipartite graphs}.
\newblock In \emph{Proceedings of the 2018 Conference of the North American
  Chapter of the Association for Computational Linguistics: Human Language
  Technologies, Volume 2 (Short Papers)}, pages 667--672, New Orleans,
  Louisiana. Association for Computational Linguistics.

\bibitem[{Chen et~al.(2018)Chen, Zhang, Wu, Yan, and
  Li}]{chen-EtAl:2018:EMNLP9}
Jun Chen, Xiaoming Zhang, Yu~Wu, Zhao Yan, and Zhoujun Li. 2018.
\newblock \href {http://www.aclweb.org/anthology/D18-1439} {Keyphrase
  generation with correlation constraints}.
\newblock In \emph{Proceedings of the 2018 Conference on Empirical Methods in
  Natural Language Processing}, pages 4057--4066, Brussels, Belgium.
  Association for Computational Linguistics.

\bibitem[{Chen et~al.(2019)Chen, Chan, Li, Bing, and
  King}]{chen-etal-2019-integrated}
Wang Chen, Hou~Pong Chan, Piji Li, Lidong Bing, and Irwin King. 2019.
\newblock \href {https://www.aclweb.org/anthology/N19-1292} {An integrated
  approach for keyphrase generation via exploring the power of retrieval and
  extraction}.
\newblock In \emph{Proceedings of the 2019 Conference of the North {A}merican
  Chapter of the Association for Computational Linguistics: Human Language
  Technologies, Volume 1 (Long and Short Papers)}, pages 2846--2856,
  Minneapolis, Minnesota. Association for Computational Linguistics.

\bibitem[{Evans and Zhai(1996)}]{evans-zhai:1996:ACL}
David~A. Evans and Chengxiang Zhai. 1996.
\newblock \href {https://doi.org/10.3115/981863.981866} {Noun phrase analysis
  in large unrestricted text for information retrieval}.
\newblock In \emph{Proceedings of the 34th Annual Meeting of the Association
  for Computational Linguistics}, pages 17--24, Santa Cruz, California, USA.
  Association for Computational Linguistics.

\bibitem[{Gu et~al.(2016)Gu, Lu, Li, and Li}]{gu:2016:ACL}
Jiatao Gu, Zhengdong Lu, Hang Li, and Victor~O.K. Li. 2016.
\newblock \href {https://doi.org/10.18653/v1/P16-1154} {Incorporating {Copying}
  {Mechanism} in {Sequence}-to-{Sequence} {Learning}}.
\newblock In \emph{Proceedings of the 54th {Annual} {Meeting} of the
  {Association} for {Computational} {Linguistics} ({Volume} 1: {Long}
  {Papers})}, pages 1631--1640, Berlin, Germany. Association for Computational
  Linguistics.

\bibitem[{Jin et~al.(2013)Jin, Kan, Ng, and He}]{jin-etal-2013-mining}
Yiping Jin, Min-Yen Kan, Jun-Ping Ng, and Xiangnan He. 2013.
\newblock \href {https://www.aclweb.org/anthology/D13-1073} {Mining scientific
  terms and their definitions: A study of the {ACL} anthology}.
\newblock In \emph{Proceedings of the 2013 Conference on Empirical Methods in
  Natural Language Processing}, pages 780--790, Seattle, Washington, USA.
  Association for Computational Linguistics.

\bibitem[{Kedzie et~al.(2018)Kedzie, McKeown, and
  Daumé~III}]{kedzie-mckeown-daumeiii:2018:EMNLP}
Chris Kedzie, Kathleen McKeown, and Hal Daumé~III. 2018.
\newblock \href {https://doi.org/10.18653/v1/D18-1208} {Content {Selection} in
  {Deep} {Learning} {Models} of {Summarization}}.
\newblock In \emph{Proceedings of the 2018 {Conference} on {Empirical}
  {Methods} in {Natural} {Language} {Processing}}, pages 1818--1828, Brussels,
  Belgium. Association for Computational Linguistics.

\bibitem[{Kim et~al.(2010)Kim, Medelyan, Kan, and
  Baldwin}]{kim-EtAl:2010:SemEval}
Su~Nam Kim, Olena Medelyan, Min-Yen Kan, and Timothy Baldwin. 2010.
\newblock \href {http://www.aclweb.org/anthology/S10-1004} {Semeval-2010 task 5
  : Automatic keyphrase extraction from scientific articles}.
\newblock In \emph{Proceedings of the 5th International Workshop on Semantic
  Evaluation}, pages 21--26, Uppsala, Sweden. Association for Computational
  Linguistics.

\bibitem[{Mao and Lu(2017)}]{Mao2017}
Yuqing Mao and Zhiyong Lu. 2017.
\newblock \href {https://doi.org/10.1186/s13326-017-0123-3} {Mesh now:
  automatic mesh indexing at pubmed scale via learning to rank}.
\newblock \emph{Journal of Biomedical Semantics}, 8(1):15.

\bibitem[{Marcu(1997)}]{marcu1997rhetorical}
Daniel Marcu. 1997.
\newblock The rhetorical parsing of unrestricted natural language texts.
\newblock In \emph{35th Annual Meeting of the Association for Computational
  Linguistics}, pages 96--103.

\bibitem[{Marujo et~al.(2012)Marujo, Gershman, Carbonell, Frederking, and
  Neto}]{MARUJO12.672}
Luís Marujo, Anatole Gershman, Jaime Carbonell, Robert Frederking, and
  JoaÌƒo~P. Neto. 2012.
\newblock Supervised topical key phrase extraction of news stories using
  crowdsourcing, light filtering and co-reference normalization.
\newblock In \emph{Proceedings of the Eight International Conference on
  Language Resources and Evaluation (LREC'12)}, Istanbul, Turkey. European
  Language Resources Association (ELRA).

\bibitem[{Meng et~al.(2017)Meng, Zhao, Han, He, Brusilovsky, and
  Chi}]{P17-1054}
Rui Meng, Sanqiang Zhao, Shuguang Han, Daqing He, Peter Brusilovsky, and
  Yu~Chi. 2017.
\newblock \href {https://doi.org/10.18653/v1/P17-1054} {Deep keyphrase
  generation}.
\newblock In \emph{Proceedings of the 55th Annual Meeting of the Association
  for Computational Linguistics (Volume 1: Long Papers)}, pages 582--592.
  Association for Computational Linguistics.

\bibitem[{Over(2001)}]{overintroduction}
Paul Over. 2001.
\newblock Introduction to duc-2001: an intrinsic evaluation of generic news
  text summarization systems.

\bibitem[{Sood et~al.(2007)Sood, Owsley, Hammond, and
  Birnbaum}]{DBLP:conf/icwsm/SoodOHB07}
Sanjay Sood, Sara Owsley, Kristian~J. Hammond, and Larry Birnbaum. 2007.
\newblock \href {http://www.icwsm.org/papers/paper10.html} {Tagassist:
  Automatic tag suggestion for blog posts}.
\newblock In \emph{Proceedings of the First International Conference on Weblogs
  and Social Media, {ICWSM} 2007, Boulder, Colorado, USA, March 26-28, 2007}.

\bibitem[{Tunkelang(2009)}]{tunkelang2009faceted}
Daniel Tunkelang. 2009.
\newblock Faceted search.
\newblock \emph{Synthesis lectures on information concepts, retrieval, and
  services}, 1(1):1--80.

\bibitem[{Wan and Xiao(2008)}]{Wan:2008:SDK:1620163.1620205}
Xiaojun Wan and Jianguo Xiao. 2008.
\newblock \href {http://dl.acm.org/citation.cfm?id=1620163.1620205} {Single
  document keyphrase extraction using neighborhood knowledge}.
\newblock In \emph{Proceedings of the 23rd National Conference on Artificial
  Intelligence - Volume 2}, AAAI'08, pages 855--860. AAAI Press.

\bibitem[{Wang et~al.(2015)Wang, Liu, and
  McDonald}]{10.1007/978-3-319-19548-3_21}
Rui Wang, Wei Liu, and Chris McDonald. 2015.
\newblock Using word embeddings to enhance keyword identification for
  scientific publications.
\newblock In \emph{Databases Theory and Applications}, pages 257--268, Cham.
  Springer International Publishing.

\bibitem[{Witten et~al.(2009)Witten, Bainbridge, and Nichols}]{witten2009build}
Ian~H Witten, David Bainbridge, and David~M Nichols. 2009.
\newblock \emph{How to build a digital library}.
\newblock Morgan Kaufmann.

\end{thebibliography}
